\documentclass[a4paper,usenatbib]{jpconf}
\usepackage{graphicx}
\newcommand{\integral}{{\it INTEGRAL}}
\newcommand{\rxte}{{\it RXTE}}

\newcommand{\beppo}{{\it BeppoSAX}}
\newcommand{\cgro}{{\it CGRO}}
\newcommand{\asca}{{\it ASCA}}
\newcommand{\granat}{{\it Granat}}
\newcommand{\cyg}{{Cyg~X-1}}
\newcommand{\gx}{{GX~339$-$4}}
\newcommand{\src}{{1E~1740.7--2942}}
\def\ergcms{erg cm$^{-2}$ s$^{-1}$ }

\def\phs{ph~cm$^{-2}$~s$^{-1}$}

\def\msun{$M_{\odot}$}

\begin{document}
\title{High-energy observations of black hole binaries with the INTEGRAL satellite}

\author{Melania Del Santo}

\address{INAF/IASF-Roma, via del Fosso del Cavaliere 100, 00133, Roma, Italy}

\ead{melania.delsanto@iasf-roma.inaf.it}

\begin{abstract}
Black-hole binaries are important sources through which studying accretion onto compact objects.
In the X/$\gamma$-ray domain, these objects show several and complex spectral behaviours and transitions.
Based on \integral\ observations collected during the last eight years, we have now a new view on the high energy emission of black-hole binary.
An additional component above 200 keV has been observed in a few systems, during either hard/intermediate or low/hard states. 
The nature of this hard-tail is still debated, as also the one observed in soft states. 
However, among a number of models, it is usually attributed to the presence of a small fraction of non-thermal electrons in a hot-Comptonising plasma. 
I review the high energy emission from black hole binary systems and report on some \integral\ observations of three different objects: \src, \gx, \cyg.
\end{abstract}

\section{Introduction}
When a compact object, like black hole (BH) or neutron star (NS),  accretes matter from a normal star, the inner part of the  accretion flow (disc) becomes a bright X-ray emitter.
In such a system, i.e. X-ray binary, matter is transferred through the inner Lagrangian point, when the  donor is less massive than the compact object (Low Mass X-ray Binary, LMXB)
or the compact object captures mass from the wind of a massive star, usually B, O, Be or supergiant (High Mass X-ray Binary, HMXB).
In both cases, the fate of the accreting matter depends on its angular momentum and physical processes by which it loses this angular momentum, 
and on the radiation processes involved (see  \cite{frank02} for a review on the accretion power in astrophysics). 

In the early 1970Õs and for the next two decades, most of the modeling
effort of accretion flows onto neutron stars and black holes was based on two
assumptions:  i) the accretion flows were assumed to be loosing angular
momentum at high rates because of an unspecified process, with the effective kinematic
viscosity typically taken to be proportional to the pressure;
ii) the radiation processes were assumed to be very efficient, so
that the resulting accretion flows were relatively cool, in the form of geometrically thin accretion disks (see  \cite{ss73}).

During the last decade, theoretical models of accretion flows onto compact objects
became increasingly more sophisticated. 
The two major developments were:  the identification of a magneto-hydrodynamic instability in differentially
rotating flows  \cite{balb91,balb98} and the discovery of new stable solutions to the
hydrodynamic equations that describe radiatively inefficient accretion flows \cite{nara94}. 
In these solutions, the electrons and ions have different and
high temperatures, the accretion flows are geometrically thick, and most of their
potential energy is not radiated away but is rather advected towards the compact
objects (Advection-Dominated Accretion Flows, ADAFs).
See \cite{psa08}  for a review on accreting neutron stars and black holes.

An important ingredient in the models of accretion flows onto compact objects is the
interaction of the flows with the objects themselves. This is the region in the accretion
flows where most of the X-ray and $\gamma$-ray radiation is produced and observed with high-energy telescopes. 
Clearly, the interaction depends on whether the collapsed object is a black hole or a neutron star.
Observational features such as pulsations and/or type-I bursts are proofs of the
existence of a surface: in these cases the nature of the compact object is definitely unveiled as being a neutron star.

\subsection{Spectral States of Black Hole Binaries}
\label{sec:state}
Black hole binaries (BHB) are known to show different spectral states in the X and $\gamma$-ray domain.
The two main spectral states are the Low/Hard State (LHS), with the high energy spectrum described 
by a cut-off power-law (typically $\Gamma$ $\sim$1.5 and $E_{cut} \sim$100 keV), and
the High Soft State (HSS) with a thermal component peaking at few keV and the 
high energy power-law much softer ($\Gamma > 2.2$) \cite{zdz00}.
Usually, this spectral variability is interpreted as due to changes in the geometry of 
the central parts of the accretion flow (see \cite{done07} for a recent review).
In the LHS the standard geometrically thin and optically thick disc \cite{ss73} would be truncated far away from the last stable orbit.
In the innermost parts, a hot accretion flow is responsible for the high energy emission,
via thermal Comptonisation of the soft photons coming from the truncated disc.  
In the HSS, the optically thick disc would extend close to the minimum stable orbit producing the dominant thermal component. 
The weak non-thermal emission at higher energy is believed to be due to up-scattering of the soft 
thermal disc emission in active coronal regions above the disc \cite{zdzg04}.
A number of intermediate states with spectral parameters in between the two canonical states are also observed.
A brief review on the radiative processes responsible for X-ray emission in hard (low) and soft
(high) spectral states of black-hole binaries is  \cite{zdzg04}.
See \cite{mcc} as more recent and exhaustive review on BH binaries.

\subsection{Black hole candidates}
  In the X and $\gamma$-ray domain, BHBs manifest themselves in a number
 of different spectral/temporal states  \cite{mcc}.
 Up to now, 18 object are dynamically confirmed black hole binaries \cite{mcc},
 while more than 30 objects are black hole candidates (BHC).
Certain characteristic X--ray properties of these established BHBs 
are often used to identify a BHC when the radial velocities of
the secondary cannot be measured (e.g., because the secondary is too
faint).  However, these signatures have been observed in a number of systems known to contain a NS primary\cite{tanakalewin}
and definitely cannot be used for the BH identification.
This is not surprising since the X--ray spectral/temporal properties
originate in an accretion flow that is expected to be fairly similar
whether the primary is a BH or a weakly magnetized NS.  

Of special importance is the mass function,
$f(M)~\equiv~P_{\rm orb}K_{2}^{3}/2\pi G~=~M_{1}$sin$^3i/(1+q)^{2}$.
The observables on the left side of the equation are the orbital
period, $P_{\rm orb}$, and the half--amplitude of the velocity curve
of the secondary, $K_{2}$.  On the right, the quantity of most
interest is the BH mass, $M_{1}$; the other parameters are
the orbital inclination angle, $i$, and the mass ratio,
$q~\equiv~M_{2}/M_{1}$, where $M_{2}$ is the mass of the secondary.
Thus a secure value of the mass function can be determined for a
quiescent X--ray nova or an HMXB by simply measuring
the radial velocity curve of the secondary star.  
An inspection of the equation for $f(M)$ shows that the value of the
mass function is the absolute minimum mass of the compact primary.
Thus, for 12 of the 18 BHBs, the very secure value of f(M) alone is
sufficient to show that the mass of the compact X--ray source is at
least 3~\msun, which is widely agreed to exceed
the maximum stable mass of a neutron star (NS) in GR \cite{rhoades,kalogera}.

\begin{figure}
\includegraphics[width=13cm]{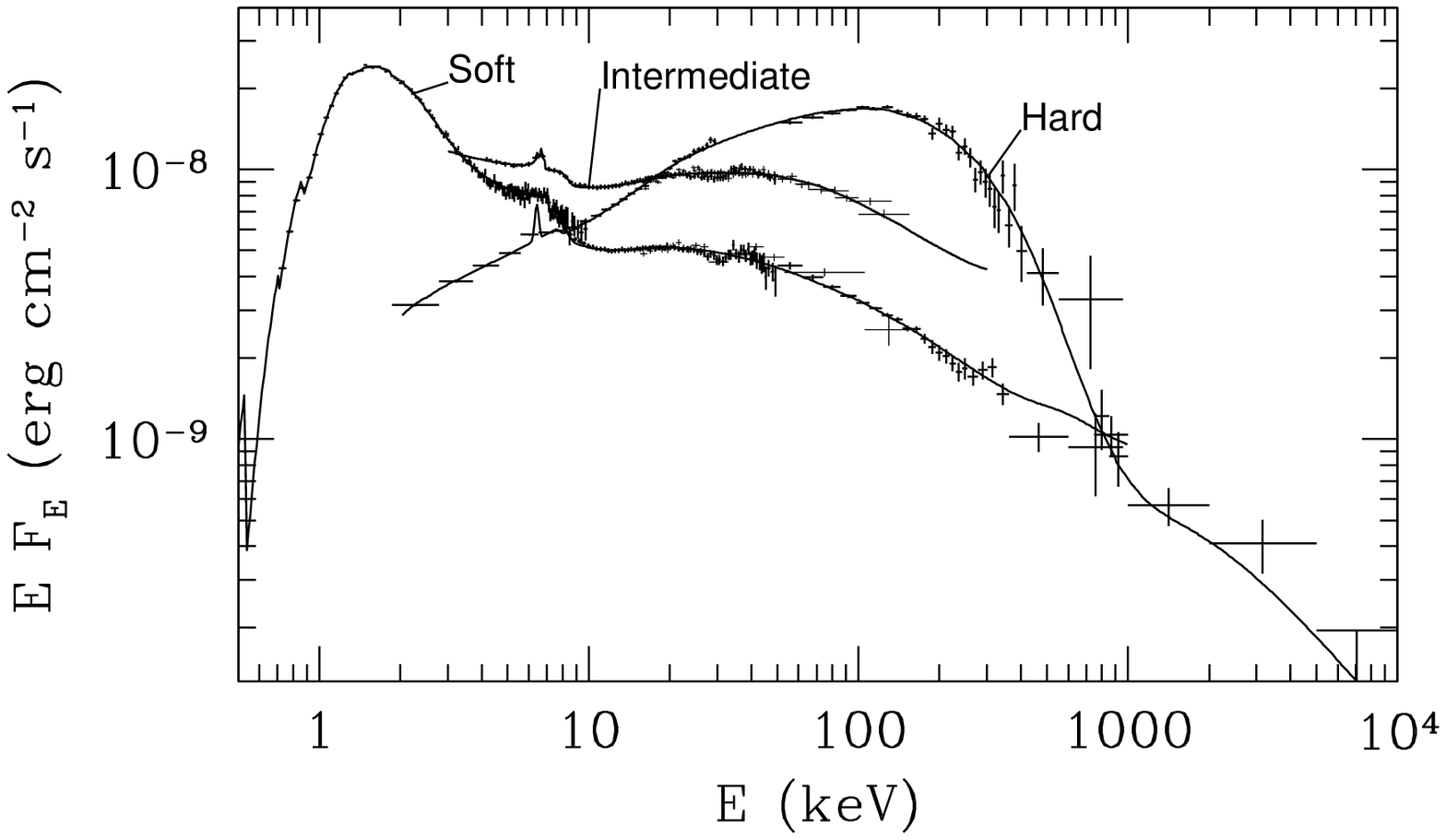}
\vskip 1cm
\centerline{\includegraphics[width=10cm]{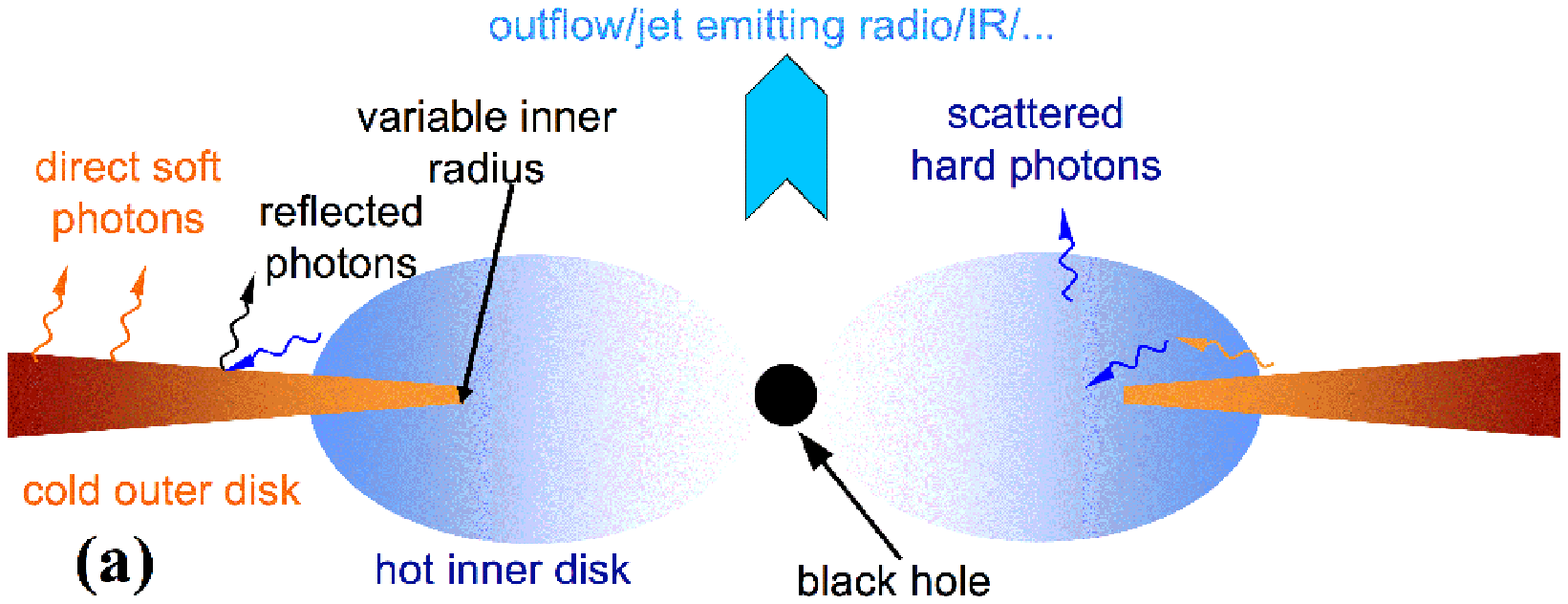}} 
\vskip 0.7cm
\centerline{\includegraphics[width=10cm]{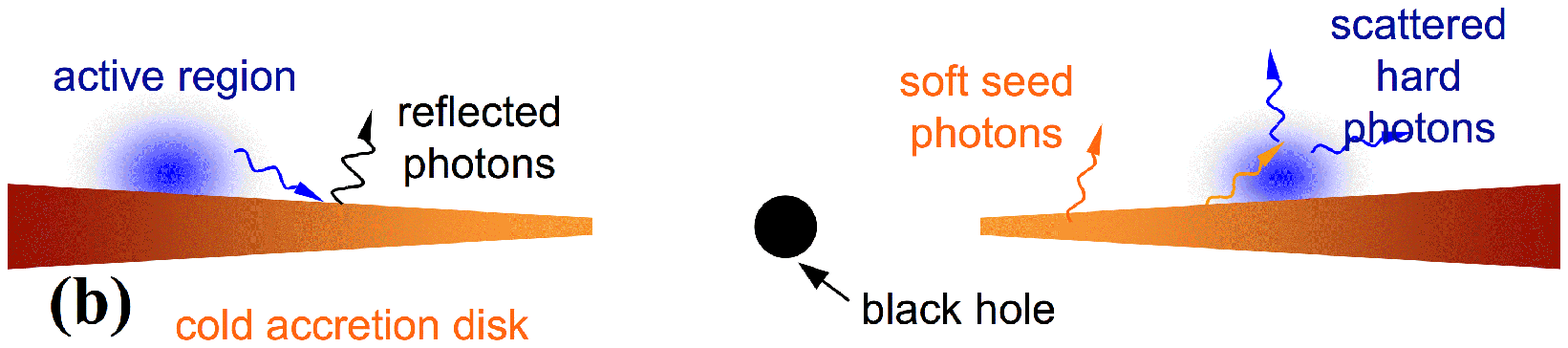}} \caption{
{\it Top}: Energy spectra in three states of \cyg\cite{zdz00}.
{\it Bottom}: (a) A schematic representation of the likely geometry in the hard state,
consisting of a hot inner accretion flow surrounded by
optically-thick accretion disk. The hot flow constitutes the base
of the jet (with the counter-jet omitted from the figure for
clarity). The disk is truncated far away from the minimum stable
orbit, but it overlaps with the hot flow. The soft photons emitted
by the disk are Compton up-scattered in the hot flow, and emission
from the hot flow is partly Compton-reflected from the disk. (b)
The likely geometry in the soft state consisting of flares/active
regions (non-thermal electrons) above an optically-thick accretion disk extending close to
the minimum stable orbit. The soft photons emitted by the disk are
Compton up-scattered in the flares, and emission from the flares is
partly Compton-reflected from the disk \cite{zdzg04}.
(Credit: Zdziarski 2000 and Zdziarski \& Gierli\'nski 2004).}
\label{geo}
\end{figure}

\begin{figure}
\centering
\includegraphics[angle=90, scale=0.5, width=9.5cm,height=5cm]{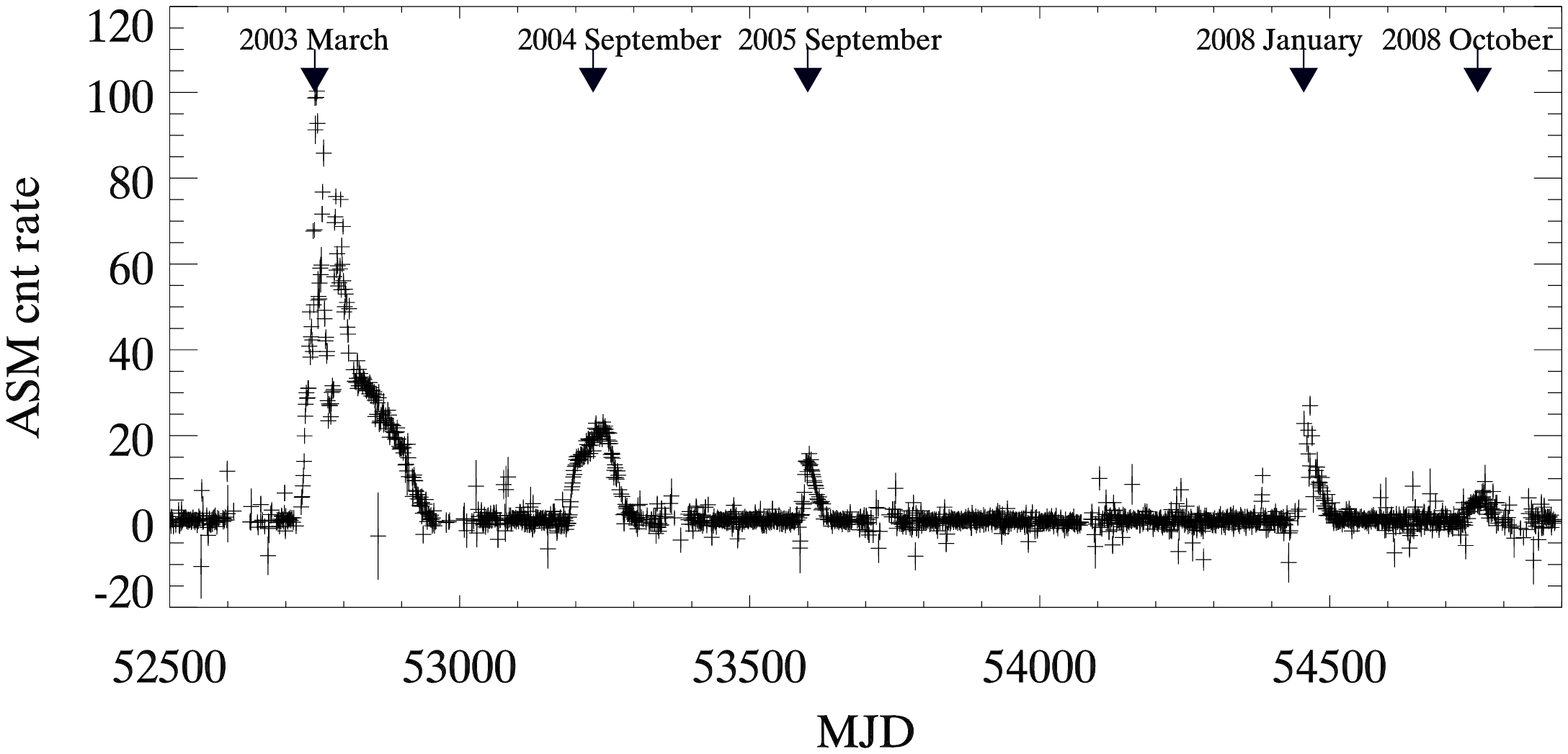}
\includegraphics[angle=0, scale=0.5,width=5.5cm]{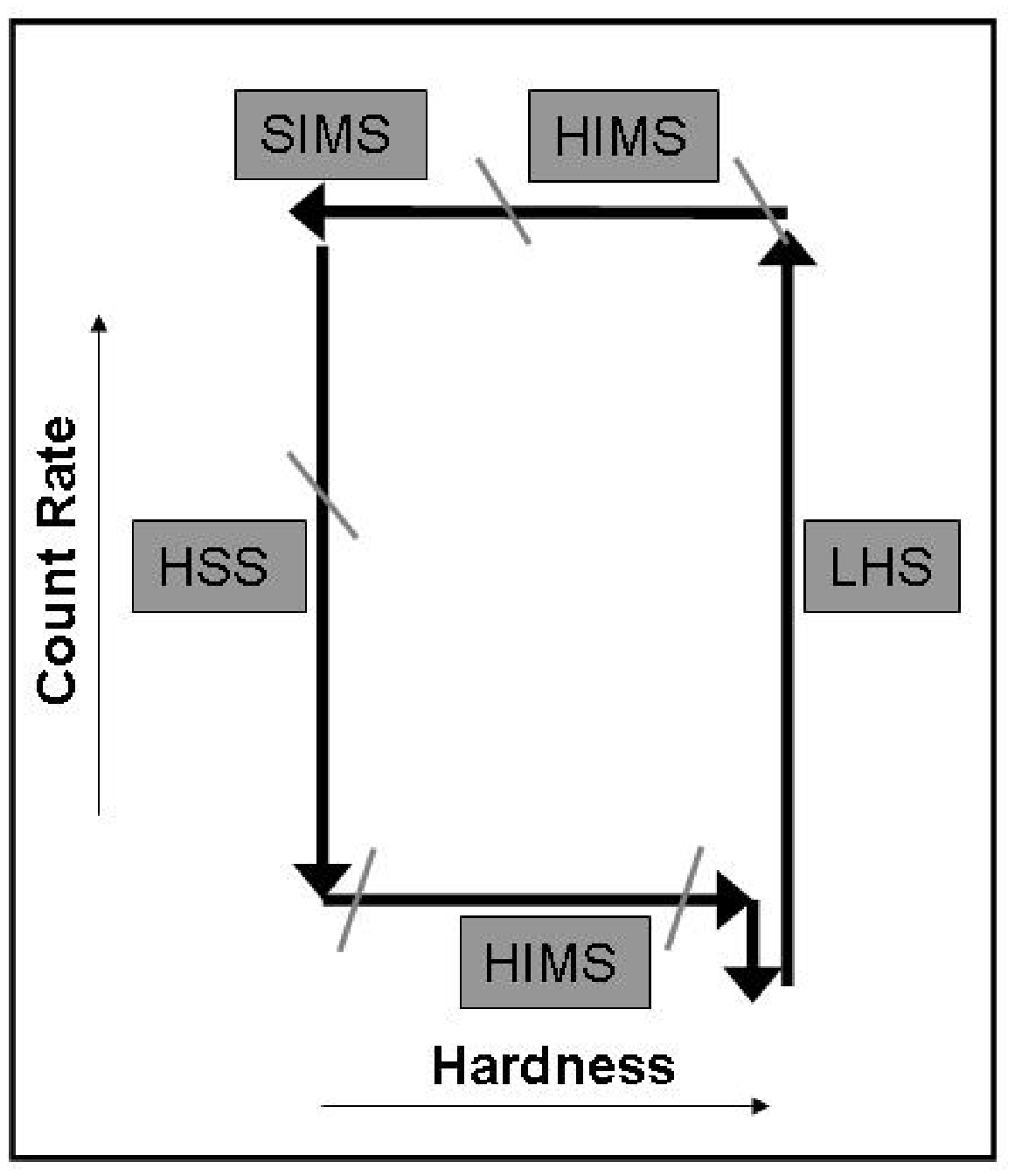}
\caption{{\it Left}: \rxte/ASM 1.5-12 keV  light curves of  H~1743--322 from 2003 until the end of 2008. Five outbursts at different flux levels have been observed by \rxte\ until 2008.
{\it Right}:  Sketch of the "q-track pattern" indicating the evolution through all the different spectral states of a typical outburst of transient  BHCs.}
\label{fig:out}
\end{figure}

\subsection{Evolution of a transient LMXB with black hole as compact object} 
During the outbursts of transient black hole candidates (BHCs), in addition to the large changes in X-ray luminosity (Fig. \ref{fig:out}), 
marked variations are observed in the properties of the timing and the energy spectrum often on very short time scales (see e.g. \cite{belloni05}). 
A BHC spends its time mostly in a quiescent state at low flux level  ($<$ 10$^{32}$--10$^{33}$ erg s$^{-1}$,  e.g. \cite{campana01}). 
When the outburst begins, the luminosity of the source increases and the X-ray spectrum is typical of the LH state (see previous section). 
The radio emission in this state indicates the presence of steady jets, while the power spectrum is dominated by a strong band limited noise ($>$30\% fractional rms). 
Then the outburst  evolves as the source increases its luminosity and its spectrum starts to change: the soft thermal component appears and becomes more important, 
the energy peak of the emission softens and the photon index of the hard component steepens ($\sim$ 2.0-2.5).
Two different states with these spectral characteristics have been defined:
 the hard intermediate state (HIMS) and the soft intermediate state (SIMS). 
 The characteristics of these two states are quite complex: the changes can be established mostly by the timing properties \cite{homan05} 
 and also by the ejection of relativistic jets associated to the transition from HIMS to SIMS \cite{fender04}.
After the SIMS, the source enters a state where the X-ray spectrum is dominated by the emission of the soft thermal component (the so called HS state). 
A non-thermal power law tail is also present without any detectable cutoff, while the power spectrum is characterised by a low-level (1-2\% fractional rms) variability.
Then the flux starts to decrease, most likely following a parallel decrease in accretion rate. 
At some point, a reverse transition is started and the path is followed backwards all the way to the LHS and then to quiescence. 
As mentioned above, the luminosity level of this back-transition is always lower than that of the corresponding forward-transition.
 The hardness-intensity diagram of transient BHCs shows such a basic pattern, the so-called  "{\it q-track}" pattern ( \cite{fender04, homan05}; see Fig. \ref{fig:out}, {\it right}).

 \section{The \integral\ satellite}
\begin{figure}
\centering
\includegraphics[width=8cm]{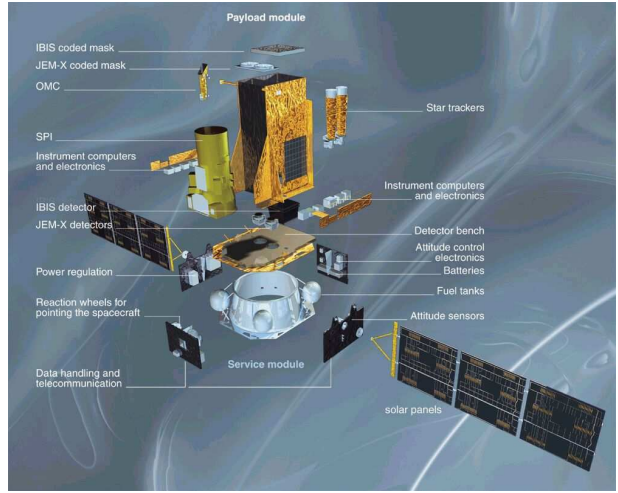}
\caption{The INTEGRAL Observatory (Courtesy of ESA).}
 \label{fig:ESA}
\end{figure}
 
The ESA \integral\ observatory  (International Gamma-Ray Astrophysics Laboratory, \cite{wink03})
was selected in June 1993 as medium-size scientific mission within ESA Horizon 2000 programme.
\integral\ is dedicated to the fine spectroscopy (2.5 keV FWHM @ 1 MeV) and imaging (angular resolution: 12' FWHM) of 
gamma-ray sources in the energy range 15 keV--10 MeV. While the
imaging is carried out by the imager IBIS\cite{ube03},
the fine spectroscopy if performed by the spectrometer SPI
\cite{vedrenne03}. Two  coaxial monitors are dedicated to the X-ray (JEM-X, 3-35 keV\cite{lund03}) and optical  (OMC\cite{mass03})
counterparts observation

SPI, IBIS and JEM-X are coded  mask telescopes. This technique is used when photons focusing 
become impossible by using standard grazing technique (see Skinner \& Connell 2004\cite{skinner04} for details).
\integral\ was launched from Baikonur on 2002, October 17$^{th}$ and inserted into an highly eccentric
orbit lasting 3 days. The data are received by the \integral\ Mission Operation Centre (MOC) in Darmstadt (Germany)
and relayed to the INTEGRAL Science Data Center (ISDC\cite{courvoisier03})
which provide the final consolidated data products to the observes
and later archived for public use. The proprietary data become
public one year after distribution to single observation PIs.

The IBIS (Imager on Board \integral\ Spacecraft) imaging instrument is optimized for survey work with a large ($29{^\circ}\times29{^\circ}$)
field of view (FOV) with excellent imaging and spectroscopy capability.
The data are collected with the low-energy detector, ISGRI (INTEGRAL Soft Gamma-Ray Imager\cite{lebrun03}),
working in the energy range 17-500 keV.
A sequence of IBIS survey catalogs has been published at
regular intervals as more data have become available \cite{bird04,bird06,bird07,bird10}.
In the last catalogue (on 2010 \cite{bird10}), 700 high-energy sources have been provided. Among them 22 are identified as black hole binaries or candidates. 
Up to now, more than 25 BHB/BHC have been observed by \integral: it is interesting to note that
above 150 keV (up to 300 keV) the point sources population is  totally dominated by these sources.

 \begin{figure}
\begin{center}
\includegraphics[height=10.5cm, angle=90]{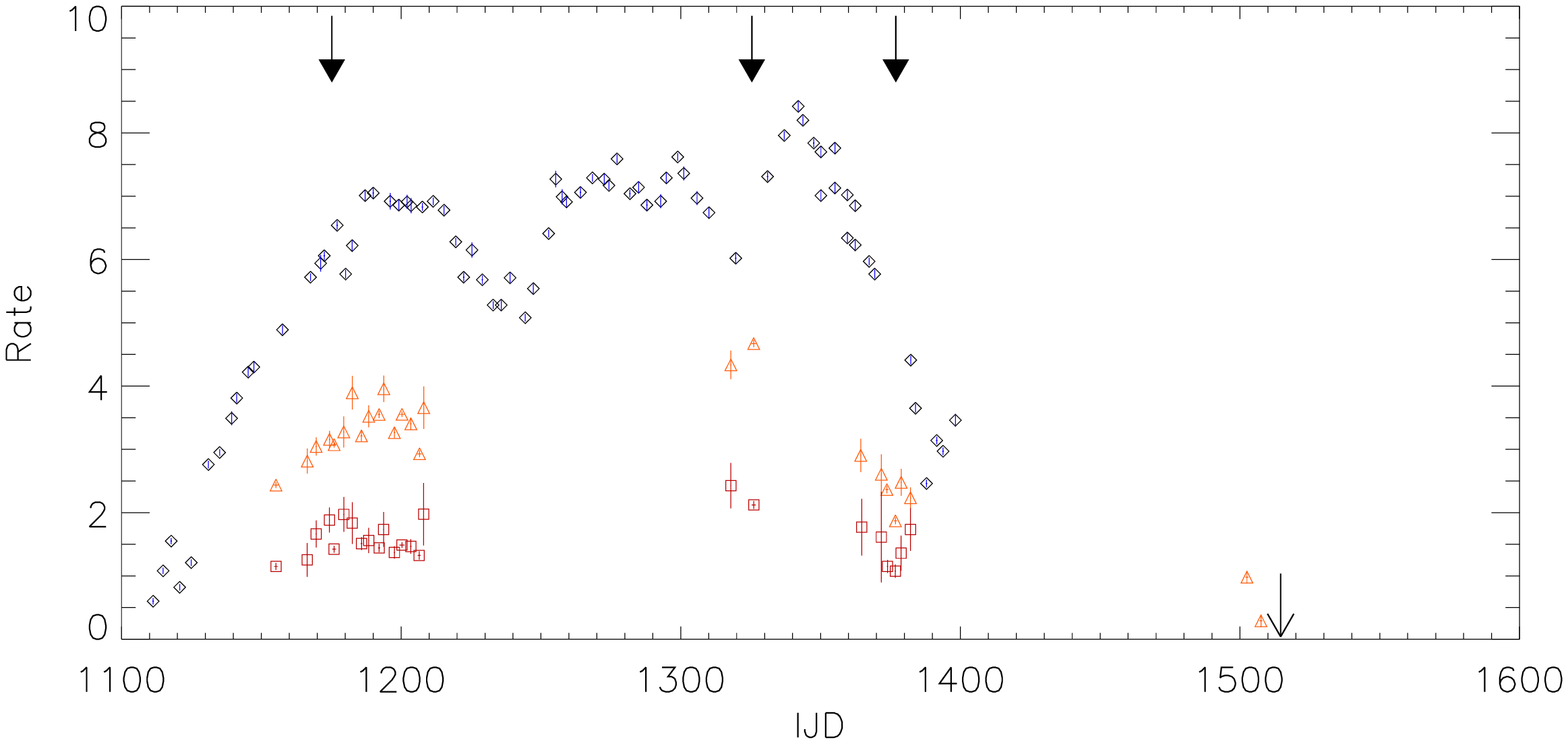}
\includegraphics[height=5cm]{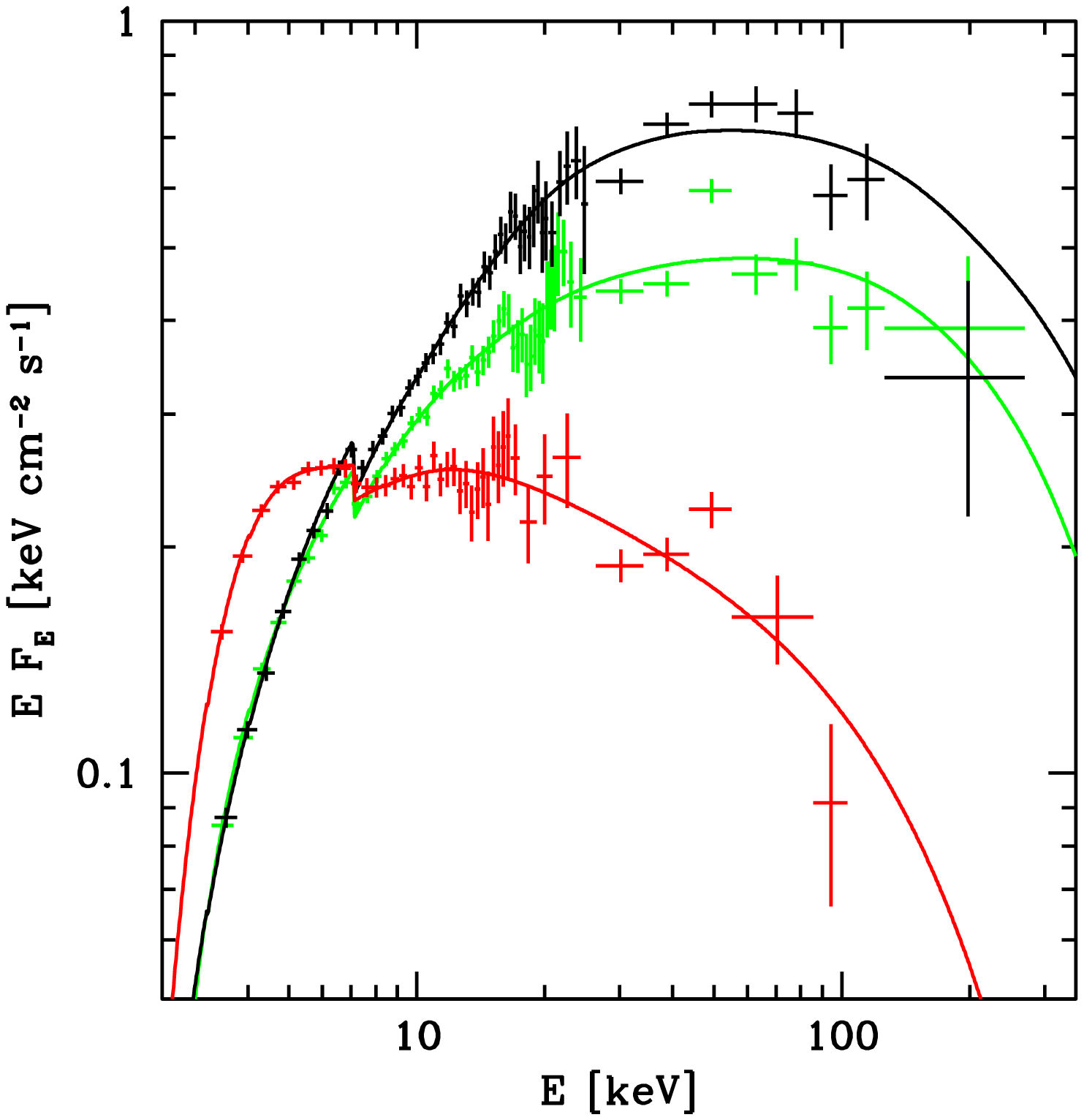}
\caption{{\it Left}: Lightcurves from PCA (8--25 keV, blue diamonds), and from IBIS/ISGRI in the two bands 30--50 keV (orange triangles)
and 50--80 keV (red squares). Solid arrows indicate the three periods used for the spectral analysis.  
Simple arrow indicates when the source started to be no longer visible by IBIS (beginning 2004). 
{\it Right}: The observed spectra, as fitted by a thermal Comptonization model \cite{delsanto05}. 
The green and  black spectra are canonical low/hard states;  the red one is a soft/intermediate state.}
\label{fig:micro} 
\end{center}
\end{figure}

\section{The microquasar 1E1740.7--2942} 
\src\ is a bright hard  X-ray source located at less than one 
degree off the Galactic Centre \cite{her84},
classified as Black Hole Candidate \cite{suni91} .
When in 1992 \cite{mira92} it was discovered a double sided radio jet reaching large angular
distances from the core ($\sim$ 1'), the "microquasar" class was born with \src\
as its first member.
\\
All observations performed so far revealed that \src\ spends most of the time in the canonical LH state of BHC.
In a few occasions, soft spectral states have been observed during \src\ low flux levels \cite{smith02}.
Simultaneous \integral\ and \rxte\ broad-band spectral study
performed in 2003  (Fig. \ref{fig:micro}, {\it left}) reported on an intermediate/soft spectral state (Fig. \ref{fig:micro}, {\it right}) 
occurred just before the source quenching \cite{delsanto05}.
One interpretation is a simple model with two simultaneous, independent accretion flows: a
thin disk and a hot halo. A drop in the accretion rate affecting both flows would propagate through the halo
immediately but might take up to several weeks to propagate through the disk. While the inner halo is thus
temporarily depleted compared to the disk, a temporary soft state is expected  \cite{smith02}.

\subsection{High-energy spectral study} 
\begin{figure}
\begin{center}
\includegraphics[height=6cm,angle=90]{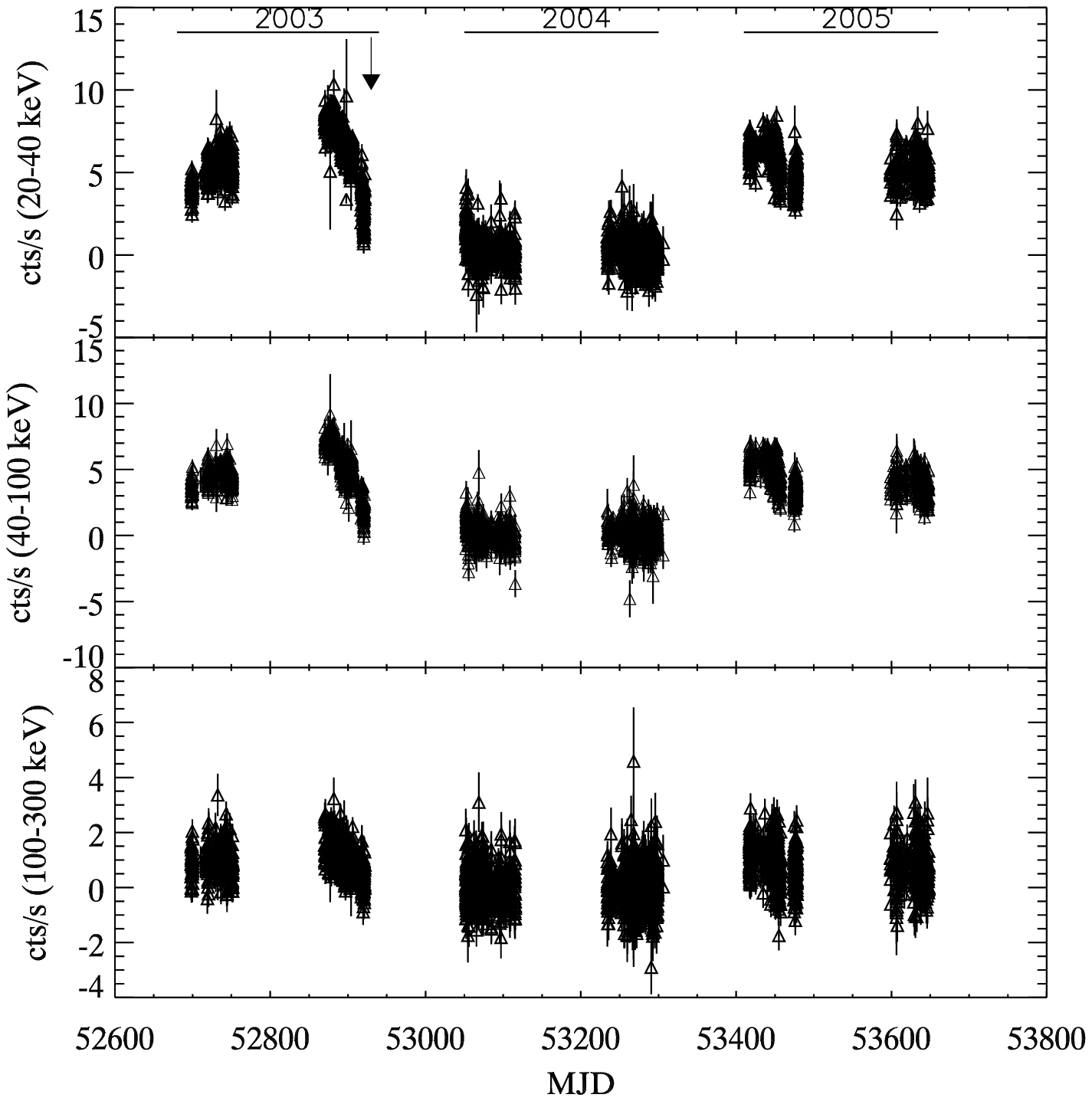}
\includegraphics[height=6cm]{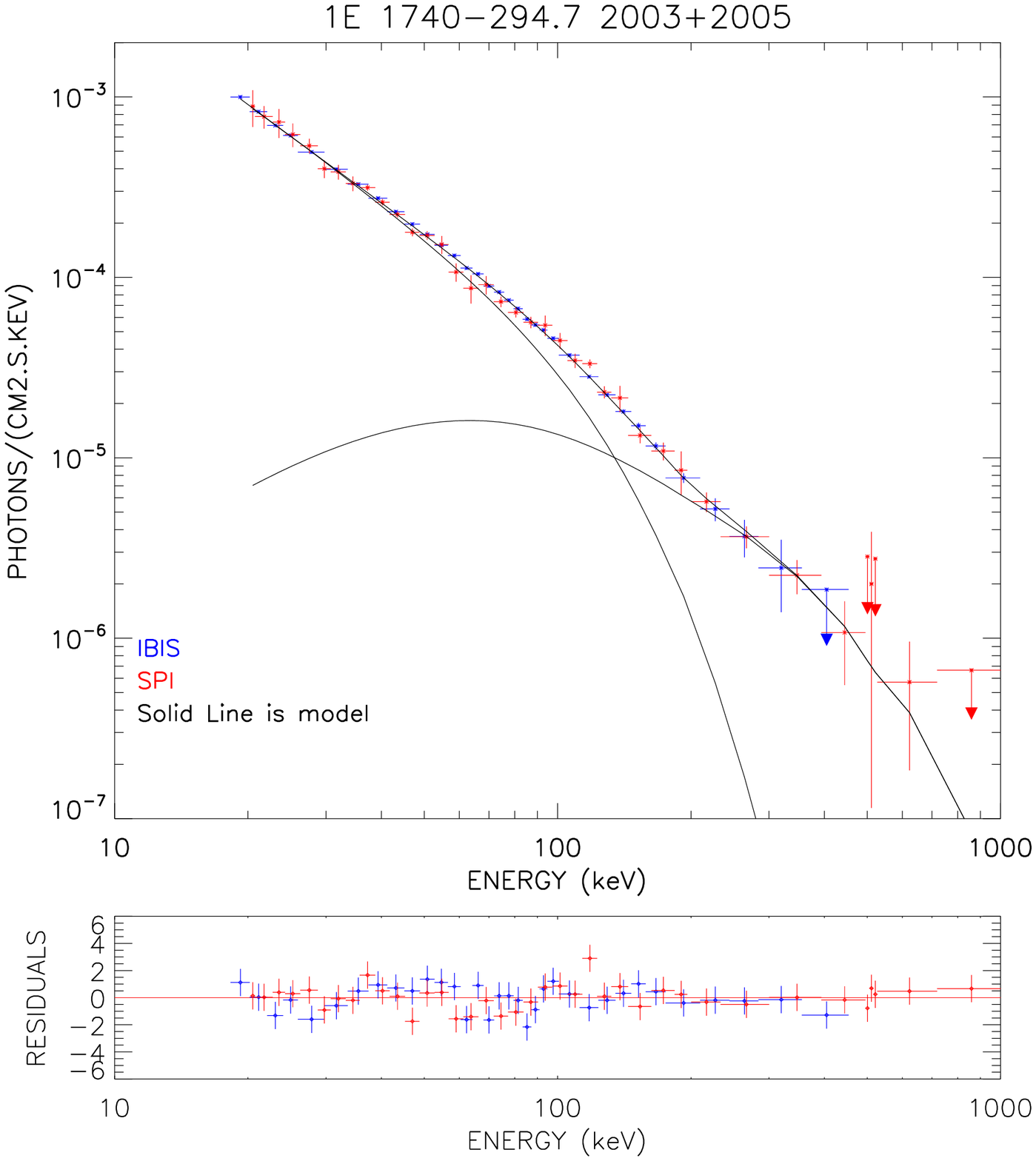}
\caption{{\it Right}: \src\ temporal behaviour observed  with IBIS/ISGRI in 2003, 2004, 2005 in 3 different energy bands.
{\it Left}: \src\ spectrum observed by IBIS and SPI for 2003 + 2005 observations.
Solid line is a two comptonisation model, with hotter temperature fixed to 100 keV. The optical depths are typical $\tau{_1}$=1.6 and $\tau{_2}$=2.2, respectively.}
\label{fig:micro2} 
\end{center}
\end{figure}

In Fig. \ref{fig:micro2} ({\it left}),  I show the  temporal behaviour of \src\ as observed with IBIS/ISGRI in 2003, 2004, 2005. 
Excluding occasional soft states, two main states have been observed during these three years: 
the canonical LH state with a mean flux of $\sim$ 50  mCrab and a "dim" state during which \src\ was at  2 mCrab flux level \cite{delsanto08a}.

We averaged all IBIS/ISGRI and SPI Low/Hard spectra (from 20 keV) collected during the whole analysed period \cite{bouchet09}. 
The averaged spectrum (roughly 2 Ms of exposure time) allowed us to achieve a better statistics at high-energy and eventually give better constraints to the spectral parameters. 
For the first time the continuum of the LH state of \src\  has been measured up to $\sim$600 keV. 
 When adjusted with a single Comptonisation model  (COMPTT \cite{tita94}),   an additional component is strongly required to fit
the data above 200 keV.  
This high energy component has been observed in several BHCs \cite{mac00, zdz01}, usually during high soft spectral states, and  explained
as  Compton up-scattering by a non-thermal electrons population \cite{zdzg04}. 
Alternatively, we show that a model composed by two thermal Comptonisations provides an equally representative description of the data: 
the temperature of the first population of electrons (kT${_1}$) results as $\sim$30 keV while the second, kT$_{2}$ is fixed at 100 keV (Fig. \ref{fig:micro2}, {\it right}). 
This two temperature model could  either correspond to  two distinct heating mecanisms/regions or
reflect the presence of a gradient of temperature in the Comptonising plasma \cite{malzac00}.

\subsection{The annihilation line} 

In 1990, the SIGMA telescope on-board \granat\ detected a broad line  
around the electron-positron annihilation energy \cite{bouchet91, suny91}. This 
transient   feature appeared clearly during a 13 hours observation 
and then possibly in two further occasions but at a less significant level.  Numerous works dedicated to
similar line searches have followed and all led to negative conclusions (see for example  \cite{cheng98}
and references therein). In this context, it has been  interesting   
to perform a deep analysis of  SPI data in this energy domain,
and to seek for  any feature around 511 keV associated with \src.
\\
The superior energy resolution of the SPI telescope allows for a specific dedicated study 
of this topic. Indeed, for the first time, an instrument is capable to look for
a narrow feature in this particular source. During the first year  of observations, no evidence for point source
emission at 511 keV has been detected with SPI.  The upper limit at 3.5 $\sigma$ level is $1.6 \times 10^{-4}$ \phs\ for
a narrow line \cite{teegar06} while the IBIS  data set a 2 $\sigma$ upper limit 
of $4.1 \times 10^{-4}$ \phs\ in the 535--585 keV energy band for 
an exposure time equal to 1.5 Ms  \cite{decesare11}.

Even though the complexity of the considered region makes
 it difficult to attribute firmly the detected emission to \src, 
the presence of photons with energy greater than several hundreds of keV
 in a more or less persistent way (something as half or 2/3 of the time),
  together with previously reported annihilation emission,  support  a scenario in which 
\src\ is a source of positrons. Indeed, as proposed by van Oss \& Belyanin (1995),
 a plasma detected with a temperature much lower than 1 MeV is able to produce
positrons through photon-photon absorption. The basic argument is that the 
hard X-ray emission comes from the regions close to the central black hole,
where the gravity field is very strong. The high local temperature is thus 
lowered, leading to an observed value far from  the relativistic domain, while pairs 
are created in the innermost disk and driven away. Annihilation outbursts could occur 
when the accretion flow intercepts the pair wind \cite{oss95}.  
Concerning the 511 keV line itself, no feature, broad or narrow, transient or persistent,
has been found by \integral\ from point sources, confirming the rare occurrence of such a phenomenon in line with numerous
different studies already performed on this topics.

\begin{figure}
\begin{center}
\includegraphics[height=8cm]{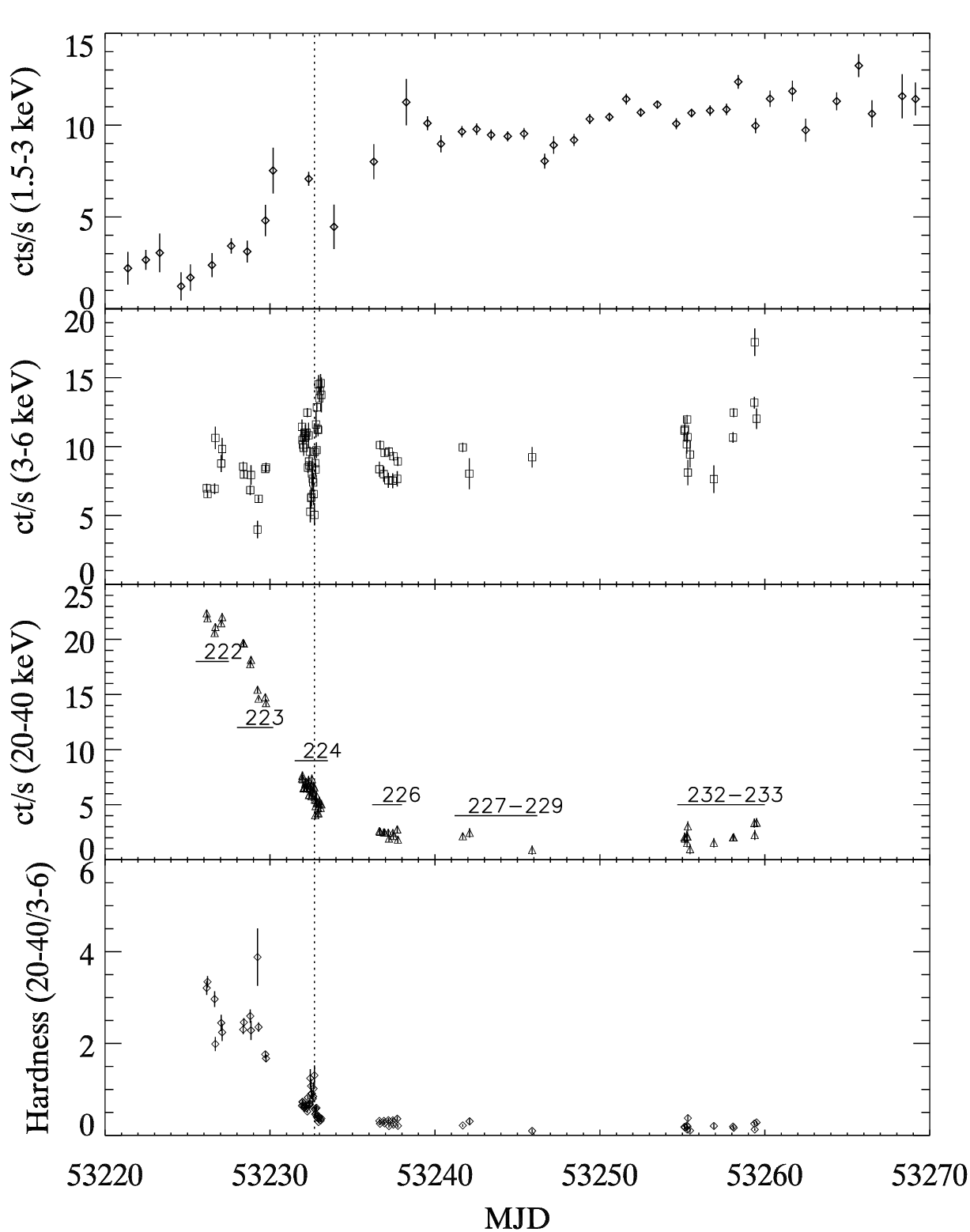}
\includegraphics[height=9.cm, width=7.cm,angle=90]{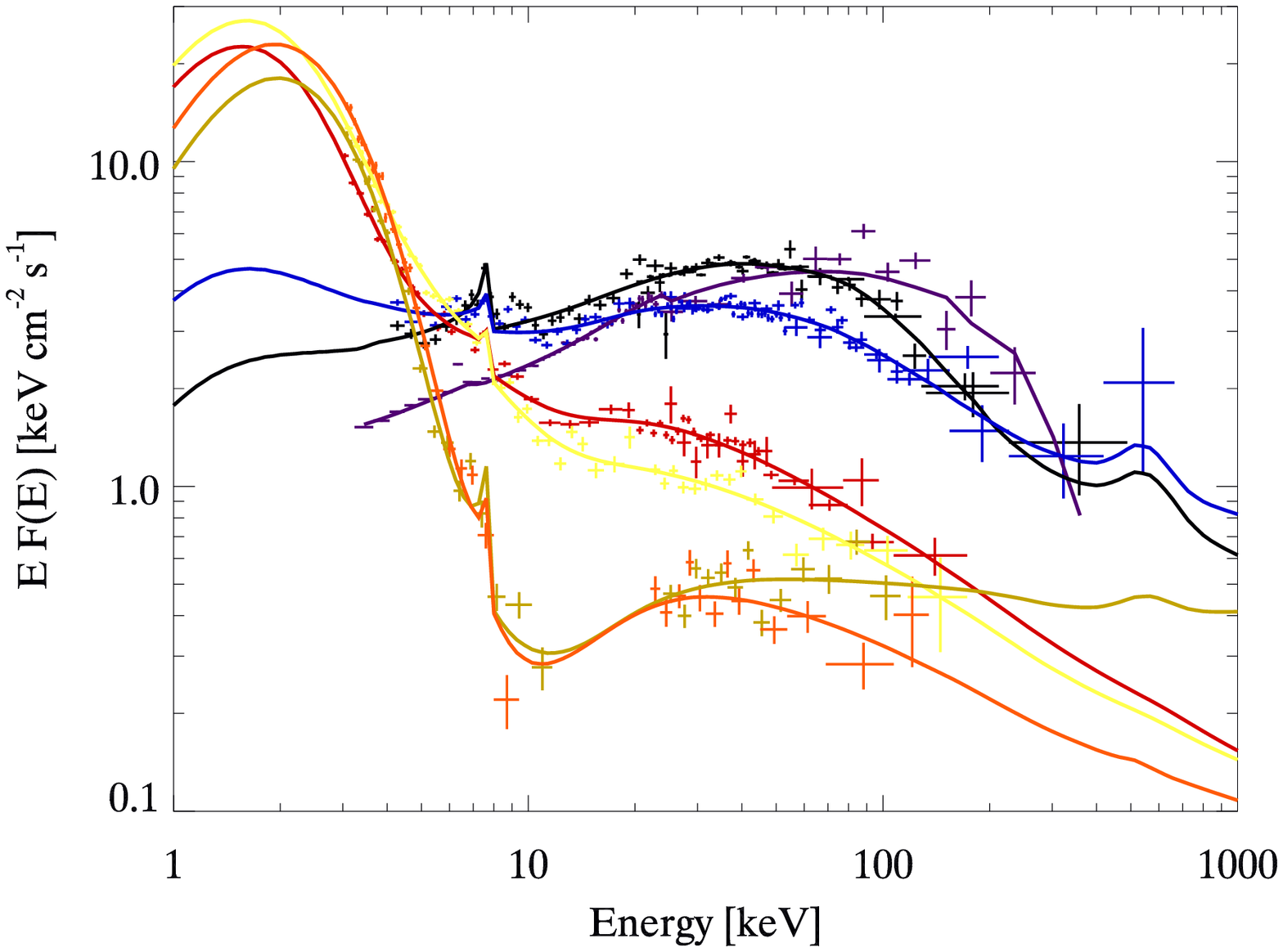}
\caption{{\it Left}:  \gx\ count rate in the energy ranges 1.5-3 keV, 3-6 keV, 20-40 keV with 
\rxte/ASM, JEM-X and IBIS/ISGRI, respectively. The related \integral\ orbits are also marked on the IBIS light curve. 
IBIS/ISGRI to JEM-X hardness ratio is shown in the bottom panel.
{\it Right}: Joint JEM-X, IBIS and SPI unfolded energy spectra (Model: {\sc eqpair}) of \gx\ averaged during different
spectral states in 2004: HIMS for spectra 1 (black), 2 (blue), 3 (red); SIMS for 4 (yellow); 
HSS for spectra 5 (green) and 7 (orange). In order to give a comparison with a pure 
Low/Hard state spectrum, we show a spectrum (violet) reported in \cite{joinet07}. }
\label{fig:gx} 
\end{center}
\end{figure}

\section{The transient black hole binary GX 339-4} 

Since its discovery \cite{markert73}, the X-ray binary \gx\ has been thoroughly  studied at all wavelength from
radio to gamma-rays. Its star companion is still unknown, even though upper limit on the optical 
luminosity allowed to classify the source as Low Mass X-ray Binary (LMXB; \cite{sha01}). 
Classified as BHC \cite{zdz98}, \gx\ is a transient source  spending long periods in outburst. 
The long-term variability of \gx\ (1987-2004) is extensively presented in  \cite{zdz04}. 
\integral\ has observed part of the 2004 outburst (August-September; Fig. \ref{fig:gx}, {\it left}) and reported on different spectral transitions \cite{delsanto08b}.

\subsection{The 2004 outburst as observed by \integral} 
In the period 2004 August 9th--September 11th \integral\  observed the \gx\ outburst  (Fig. \ref{fig:gx}, {\it left}) during different spectral states,
namely the hard/intermediate, soft/intermediate and the high/soft states. 
Joint JEM-X, IBIS and SPI collected during the different states of \gx\ have been fitted {\sc eqpair} \cite{coppi99} (Fig. \ref{fig:gx}, {\it right} and Tab. \ref{tab:fit}).
In the {\sc eqpair} model, emission of the disc/corona system is modeled by a
spherical hot plasma cloud with continuous acceleration of electrons illuminated by soft
photons emitted by the accretion disc. 
At high energies the distribution of electrons is non-thermal,
while at low energies a Maxwellian distribution with temperature $kT_{e}$ is established.
Our detailed spectral analysis of time averaged spectra at different stages of the transition confirms that 
the spectral transition in \gx\ is driven by changes in the soft cooling photon flux in the corona associated 
with an increase of disc temperatures (leading to dramatic increase of disc luminosity).
The measured disc temperature versus luminosity relation suggests that the internal disc radius decreases \cite{delsanto08b}.
Although other models such as dynamic accretion disc corona models cannot be ruled out, 
these results are consistent with the so-called truncated disc model \cite{done07}.

\begin{table}
\begin{center}
\caption{Best-fit parameters of the joint IBIS/ISGRI, JEM-X and SPI spectra (only rev 222, 223 and 224.1 for the spectrometer).
Fits have been performed simultaneously with {\sc eqpair} combined with {\sc diskline}. 
See Del Santo et al. (2008b) for detailed discussion.}\label{tab:fit}
\tiny
\renewcommand{\arraystretch}{1.7}
\begin{tabular}{cccccccccccc} 
\hline
\hline
Period & Rev   &  $\rm l_h/l_s$ & $\rm l_{nth}/l_{h}$&$\tau _{es}$& kT$_{bb}$ &$\Omega$/${2\pi}$& $G_{inj}$&$\chi^2_{\nu}$(dof) & 
\multicolumn{3}{c}{\rm Flux $\times 10^{-9}$} \\	
      &        &                   &            &        &     [eV]             &         &                 &   & \multicolumn{3}{c}{[\ergcms]} \\
      &        &                   &            &        &                      &         &                     &   & $Bol$ & $bb$& $Compt$  \\
\hline
1 & 222&$4.4^{+1.0}_{-0.4}$ & $0.7^{+0.2}_{-0.1}$ & $2.6^{+0.2}_{-0.9}$ & (300) & $0.2^{+0.3}_{-0.1}$ & $2.7^{+0.5}_{-0.6}$ &  0.99(243)&
11.3 & 1.0 & 9.4 \\   
2& 223&$2.1 \pm 0.4$ &$1.0^{+0.0}_{-0.2}$  & $2.0^{+0.3}_{-0.5}$  & (300) & $0.6 \pm 0.3$   & $2.6^{+0.3}_{-0.4}$  & 1.17(219)&
12.4  & 2.1 & 8.9 \\
3&224.1&$0.20^{+0.05}_{-0.07}$ & $0.71^{+0.18}_{-0.09}$ & $0.18^{+0.12}_{-0.03}$ & $380^{+54}_{-57}$& $1.0^{+0.1}_{-0.3}$& $2.9^{+0.1}_{-0.2}$ & 1.14(215)&
21.0  & 15.4 & 5.6 \\
4&224.2&$0.17^{+0.03}_{-0.02}$ & $0.6 \pm 0.1$ &$0.4^{+0.3}_{-0.1}$&  $388^{+76}_{-83}$ & (1) & $2.9^{+0.4}_{-0.2}$ & 1.06(173)&
 24.3 & 17.8   & 6.5  \\
5&226&$0.15^{+0.13}_{-0.05} $ & $0.94^{+0.06}_{-0.10} $ &  $< 0.1$     & $495^{+27}_{-43}$  &  (1) &$1.9 \pm 0.3 $  & 1.15(121)&
13.8 & 10.9  & 2.9  \\
7&232-233&$0.07^{+0.04}_{-0.01}$ &$1.0^{+0.0}_{-0.2} $  &  $ < 0.7$ & $478^{+66}_{-83}$ & (1) & $2.8 \pm 0.5$   &   1.0(98)&
18.4  & 13.8  & 4.2  \\
\hline
\end{tabular}
\vspace*{-0.1 cm} 
\end{center}
\end{table}

\subsection{The high energy excess}
During the Low/Hard state of the \gx\ outburst occurred in 1991, a high energy excess (above 200 keV) 
was observed with OSSE \cite{john93, w02}. 
In March 2004, i.e. during the increasing phase of the 2004 outburst, Joinet et al. (2007) 
observed a similar feature with SPI when \gx\ was in its canonical LH state.

The non-negligible values of $\rm l_{nth}/l_{h}$  (compactness ratio between the electron acceleration 
and total power supplied to the plasma) found also in the hard/intermediate state (spectra 1, 2  in Tab. \ref{tab:fit})
indicated a non-thermal emission being requested by the data.
Indeed, using a simple thermal Comptonisation model, some residuals are present at high energy, 
especially in the spectrum collected during the revolution 223 \cite{delsanto08b}. 
Adding a power-law with $\Gamma \sim 2$ to the pure thermal Comptonisation model,
it led to an improvement of the $\chi^2$ (see \cite{delsanto08b} for a complete discussion).

Moreover, the SPI observations showed that during the bright hard state of the 2007 outburst, 
the highest energy emission ($>$150 keV) of \gx\ was variable  \cite{drou10}. While the spectral shape at lower
energies (4-150 keV) remained more or less constant, Droulans et al. (2010)
detected the significant appearance/disappearance of a high-energy
tail. The strength of this hard tail, varying on a timescale
of less than 7 hr, is found to be positively correlated with the
total X-ray luminosity of the source. These authors interpreted the spectral
variability as triggered by only the variations of the
magnetic field in the hot electron plasma, implying a subsequent variation of the maximum
energy of the accelerated particles  \cite{drou10}.

\section{The prototype of Galactic black hole binary: Cygnus X--1}
Cygnus X--1,  the archetype of the Galactic black holes discovered in 1964s \cite{bowyer65}, is a persistent X-ray binary with a supergiant star companion.
\cyg\ is one of the best studied black hole binaries (BHBs) and it has served as an accretion disc laboratory since the end of the 1960s.
In the LH state of \cyg, the electron temperature  and the Thomson optical depth of the Comptonising plasma were found to be typically 
 $kT_{e} = 100 $ keV and  $\tau = 1$  \cite{zdzg04}. 
 Although the scattering plasma is predominantly thermal in the hard state, since 1994 there are indications that the electron distribution has some high-energy tail  i.e., it is hybrid, thermal/nonthermal. 
This evidence was firstly provided by \cgro/COMPTEL observations of \cyg\ in the hard state at  $E > 500$ keV\cite{mac94}.
The X/$\gamma$-ray HS state spectrum has been studied extensively
by simultaneous observations with \asca, \rxte, \beppo, and \cgro\ during summer 1996\cite{disalvo01,frontera01}.
In addition to the dominating black-body, a long power-law like tail extending up
to 10 MeV was discovered \cite{mac02}. 

As one of the brighter galactic hard X-ray source, \cyg\ is a prime target for the \integral\ mission.
It was extensively observed during the Performance Verification (PV) Phase of the mission, when the source was in the HSS\cite{bazzano03,bouchet03,pot03}. 
The on-board instruments offer an unprecedented simultaneous broad-band spectral coverage,  ranging from 3 keV to several MeV. 
Thus, up to now a big amount of observing time (open time and core programme for a total of about 7 Ms) have been dedicated to this target by \integral.
A number of studies of the intermediate spectral state have been performed \cite{malzac06, bel06};
a small X-ray flare during the hard state have been observed  to be coincident with the TeV emission  detected by MAGIC \cite{malzac08}.
Recently, Laurent et al. (2011) used the IBIS telescope in Compton mode and measured the polarization of the gamma-ray emission from \cyg.
While the 250-400 keV spectrum is consistent with emission dominated by Compton scattering by thermal electrons and are weakly polarized (Pf $<$20\%),
the second spectral component seen in the 400 keV--2 MeV band is strongly polarized (Pf $=$ 67$\pm$ 30\%).
The authors proposed that the MeV emission is probably related to the jet  detected in the radio band.

However, Jourdain et al. (2012) report on  \cyg\ observations performed by the SPI telescope 
distributed over more than 6 years. They investigate the hard X-rays domain, and more particularly up to the MeV region.
They found that the \cyg\ emission decreases sharply above 700 keV,
with flux values above 1 MeV (or upper limits) well below to the one reported  in\cite{laurent11}, 
while compatible with the MeV emission detected some years ago by \cgro/COMPTEL\cite{mac02}.

\begin{figure}
\centering
\includegraphics[height=6cm,angle=+90]{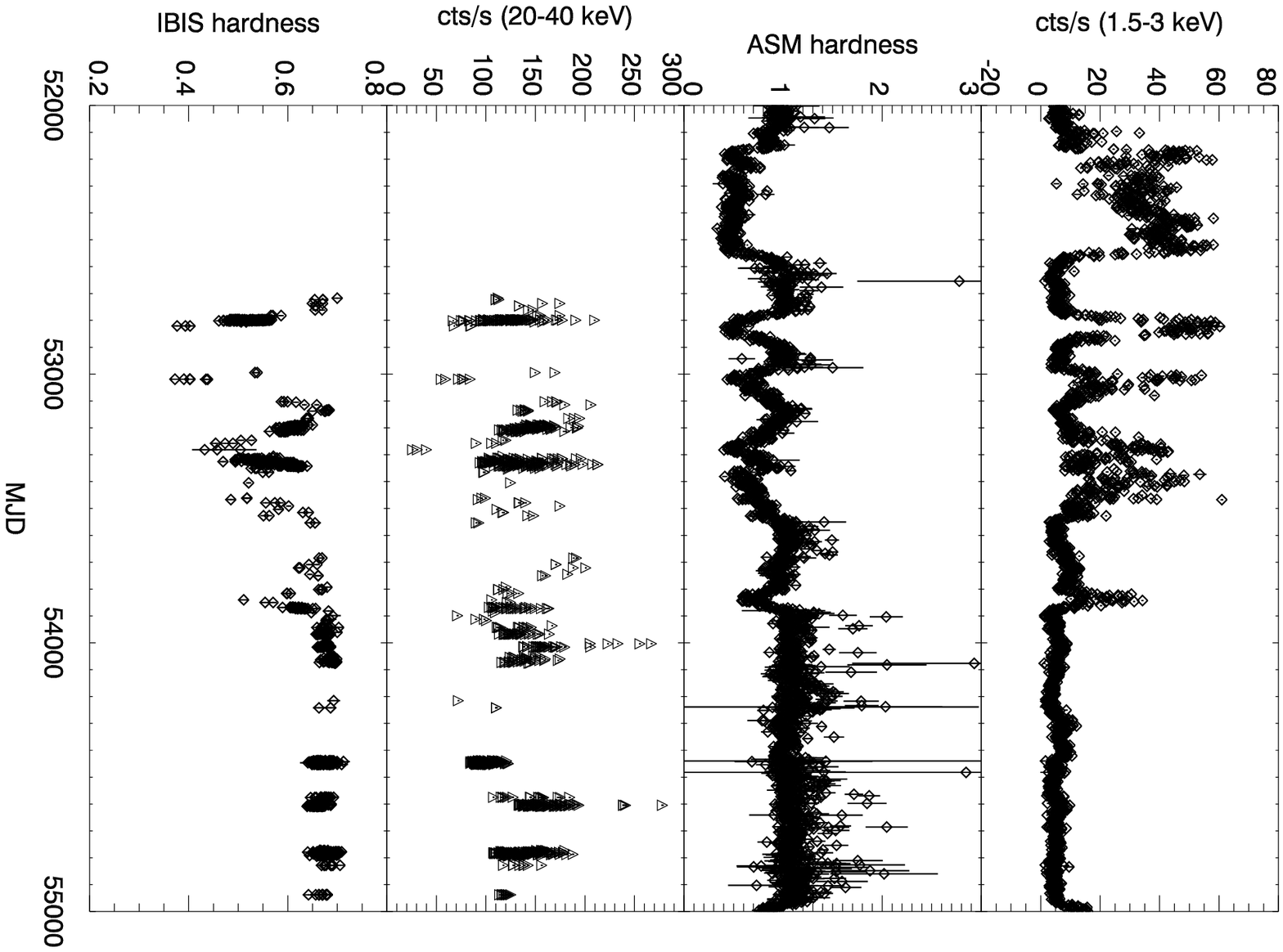}
\includegraphics[width=9.5cm]{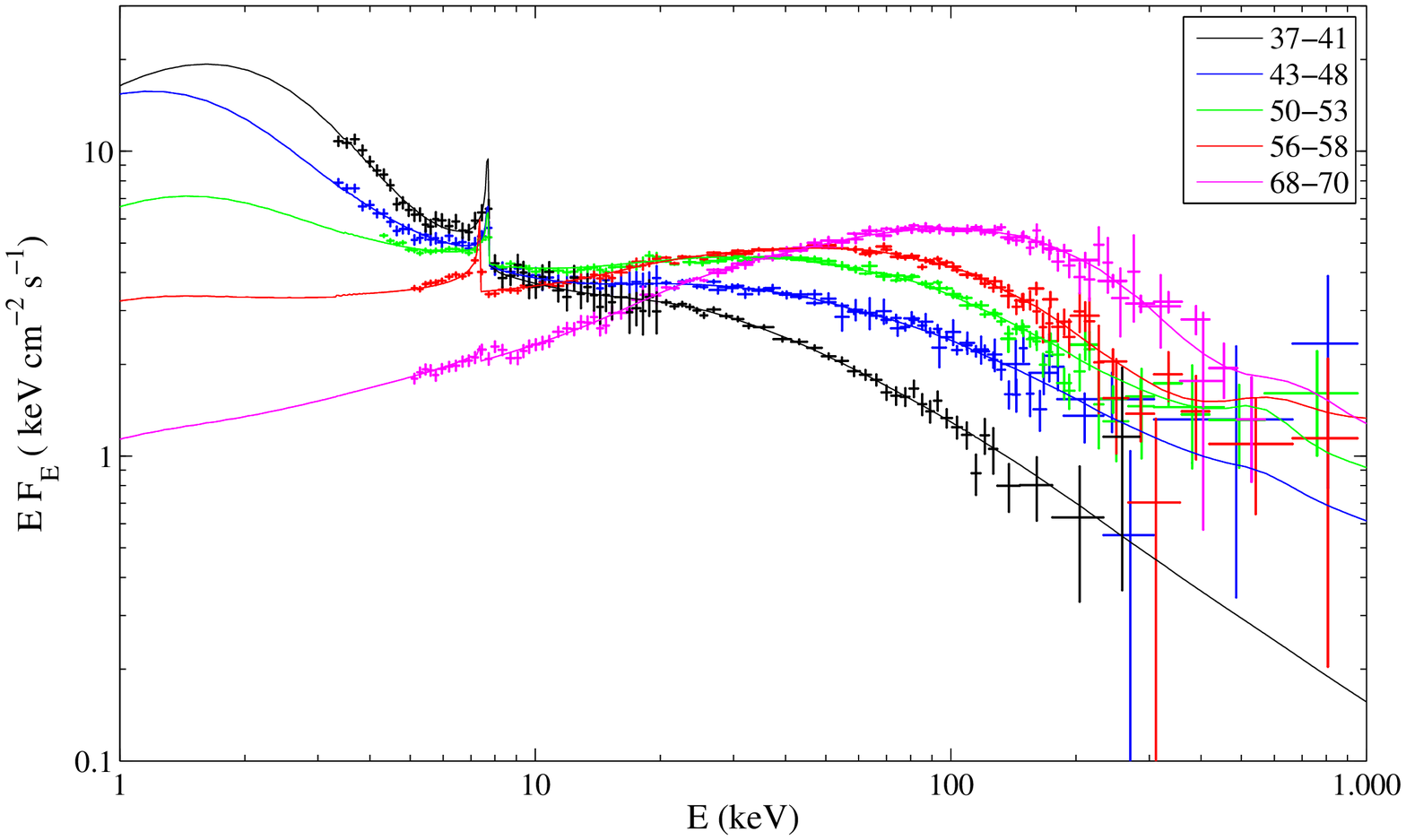}
\caption{{\it Left}: From top to bottom: ASM count rate in 1.5-3 keV, ASM hardness ratio (3-6/1.5-3 keV),  IBIS/ISGRI count rate in 20-40 keV,  IBIS/ISGRI hardness (40-80/20-40 keV).
{\it Right}: Joint JEM-X, IBIS and SPI energy spectra of \cyg\ collected in different spectral states.
The data are fitted with model {\sc eqpair} plus {\sc diskline}. Spectral parameters will be presented in Del Santo et al. (2012).
}
 \label{fig:cyg}
\end{figure}

\subsection{Long-Term Spectral Variability of Cygnus X--1 with \integral}

We have analysed six years of  Cygnus X--1  observations performed by \integral\ since March 2003 until April 2009  with the three telescopes IBIS, SPI and JEM-X.
In Fig. \ref{fig:cyg} ({\it left}), panels 3 and 4 (from top to bottom), we show the IBIS light curve and the hardness ratio of \cyg\  in the hard X-ray domain.
As a comparison at the soft X-rays, we plot the \rxte/ASM long term count rate, as well as the hardness ratio (Fig. \ref{fig:cyg} ({\it left}), panels 1 and 2).
During the first three years of the \integral\ monitoring, \cyg\ showed a very variable activity in term of both flux variation and spectral transition.
Since middle of 2006, the source entered  in an almost steady hard state, characterized by flux variability not combined with simultaneous spectral variability (Fig. \ref{fig:cyg}, {\it left}).
The work is in progress and will be presented in Del Santo et al. (2012).
In that paper, we will report on the long term behaviour  of \cyg\ using the whole \integral\ data base available until 2009.
A broad-band spectral variability study by using two different models over a wide energy band (from 3 keV up to 1 MeV) spectra has been performed.
In Fig.  \ref{fig:cyg} ({\it right}) I show preliminary \cyg\ energy spectra fitted by the hybrid thermal/non-thermal Comptonisation plus a disc reflection
model, namely {\sc eqpair} \cite{coppi99}, and a relativistic iron line line emission.

\section{Acknowledgments}
I acknowledge the following funding:
the agreement ASI-INAF I/009/10/0,
 PRIN-INAF 2009 (PI: L. Sidoli), IAPS and Universit\'e Toulouse III.
I thank a number of collaborators, in particular J. Malzac (CNRS/IRAP).
This work is based on observations with INTEGRAL, an ESA project with instruments
and Science Data Centre funded by ESA member states (especially the PI
countries: Denmark, France, Germany, Italy, Switzerland, Spain), Czech
Republic, and Poland, and with the participation of Russia and the USA.

\section{References}

\smallskip

\end{document}